\documentclass{appolb}
\usepackage{graphicx}

\newcommand{\bb}{\mbox{\boldmath $b$}}

\newcommand{\bba}{\mbox{\boldmath $b_{1}$}}
\newcommand{\bbb}{\mbox{\boldmath $b_{2}$}}
\newcommand{\bbi}{\mbox{\boldmath $b_{i}$}}

\begin{document}
\title{Production of $p\bar{p}$ pairs in UPC at the LHC%
\thanks{Presented at the XIII Workshop on Particle Correlations and Femtoscopy, 22-26 May 2018, Krak\'ow, Poland}%
}
\author{Piotr Lebiedowicz, Mariola K{\l}usek-Gawenda, Antoni Szczurek
\address{Institute of Nuclear Physics Polish Academy of Sciences,\\
ul. Radzikowskiego 152, PL-31-342 Krak\'ow, Poland}
\\
\vspace{0.5cm}
{Otto Nachtmann
\address{Institut f\"ur Theoretische Physik, Universit\"at Heidelberg,\\
Philosophenweg 16, D-69120 Heidelberg, Germany}
}
%the Name(s) of other Author(s)
%\address{affiliation}
}
\maketitle
\begin{abstract}
We discuss production of $p\bar{p}$ pairs in two-photon interactions in heavy-ion collisions.
We present predictions for the ultraperipheral, ultrarelativistic, heavy-ion collisions (UPC)
$^{208}\!Pb \, ^{208}\!Pb \to$ $^{208}\!Pb \, ^{208}\!Pb \,p \bar{p}$.
The parameters of vertex form factors are adjusted to the Belle data
for the $\gamma \gamma \to p \bar{p}$ reaction. 
To described the Belle data we include the proton-exchange, 
the $f_2(1270)$ and $f_2(1950)$ $s$-channel exchanges, as well as the hand-bag mechanism. 
Then, the total cross section and several differential distributions 
for experimental cuts corresponding to the LHC experiments are presented.
The distribution in $\mathrm{y}_{diff}$, 
the rapidity distance between the proton and antiproton, is particularly interesting.
We find the total cross sections: 100~$\mu$b for the ALICE cuts,
160~$\mu$b for the ATLAS cuts, 500~$\mu$b for the CMS cuts,
and 104~$\mu$b taking into account the LHCb cuts.
This opens a possibility to study the $\gamma \gamma \to p \bar{p}$ process in UPC at the LHC.
\end{abstract}
  
\section{Introduction}

It was presented in Ref.~\cite{Klusek-Gawenda:2017lgt} that
the ultraperipheral collisions (UPC) of heavy ions may provide 
new information on $\gamma \gamma \to p \bar{p}$ interactions
compared to the presently available data from $e^+ e^-$ collisions.
The baryon pair production via $\gamma \gamma$ fusion was measured at
electron-positron colliders by various experimental groups: 
CLEO \cite{Artuso:1993xk} at CESR,
VENUS \cite{Hamasaki:1997cy} at TRISTAN, 
OPAL \cite{Abbiendi:2002bxa} and L3 \cite{Achard:2003jc} at LEP, 
and Belle \cite{Kuo:2005nr} at KEKB.

%QCD predictions for $\gamma \gamma \to p \bar{p}$
%were first calculated in \cite{Farrar:1985gv,Farrar:1988vz} 
%using the leading twist nucleon wave functions determined 
%from QCD sum rules, see e.g. \cite{Chernyak:1984bm}. 
The calculated cross sections from the leading-twist QCD terms
\cite{Chernyak:1984bm,Farrar:1988vz}
turned out to be about one order of magnitude smaller than the experimental data
on $\gamma \gamma \to p \bar{p}$ process.
In order to explain these discrepancies,
various phenomenological approaches were suggested,
%For example, in the diquark model, 
%which is a variant of the leading-twist approach,
see e.g. \cite{Berger:2002vc} and references therein.
%the proton was considered to be a quark-diquark system
%and a diquark form factor was introduced.
In the hand-bag approach, see e.g. \cite{Diehl:2002yh},
the $\gamma \gamma \to p \bar{p}$ amplitude
was factorized into a hard $\gamma \gamma \to q \bar{q}$
subprocess and form factors describing a soft $q \bar{q} \to p \bar{p}$ transition. 
%The transition form factors could not be calculated from first
%principles in QCD and were, therefore, determined phenomenologically.
The pQCD-inspired phenomenological models have more chances to describe 
the absolute size of the cross section for $W_{\gamma\gamma} > 2.5$~GeV,
however, they contain a number of free parameters that are fitted to data.
%Moreover, most data were taken at energies which are rather low for
%the kinematic requirements of large $s$, $|t|$, $|u|$
%in the hand-bag approach.
The low $W_{\gamma\gamma}$ region of $\gamma \gamma \to p \bar{p}$
may be dominated by $s$-channel resonance contributions.
One of the effective approaches used for this region 
is the Veneziano model \cite{Odagiri:2004mn}.
While a reasonable $\sigma (W_{\gamma\gamma})$ dependence was obtained
without adjustable parameters in \cite{Odagiri:2004mn}, 
the agreement of the model with the angular distributions was only qualitative.

In our approach, described in detail in \cite{Klusek-Gawenda:2017lgt}, 
we consider all important theory ingredients
in order to achieve a quantitative description of the Belle data \cite{Kuo:2005nr}
both the dependence of the total cross section on $W_{\gamma\gamma}$
as well as corresponding angular distributions.
Then we presented predictions for the production of $p \bar{p}$ pairs 
in the ultraperipheral, ultrarelativistic, heavy-ion collisions at the LHC.

Central exclusive diffractive production of the $p\bar{p}$ pairs 
was also studied recently in proton-proton collisions \cite{Lebiedowicz:2018sdt}.

\section{Formalism}

%--------------------------------------------------------
\begin{figure}[htb]
(a)\includegraphics[width=5cm]{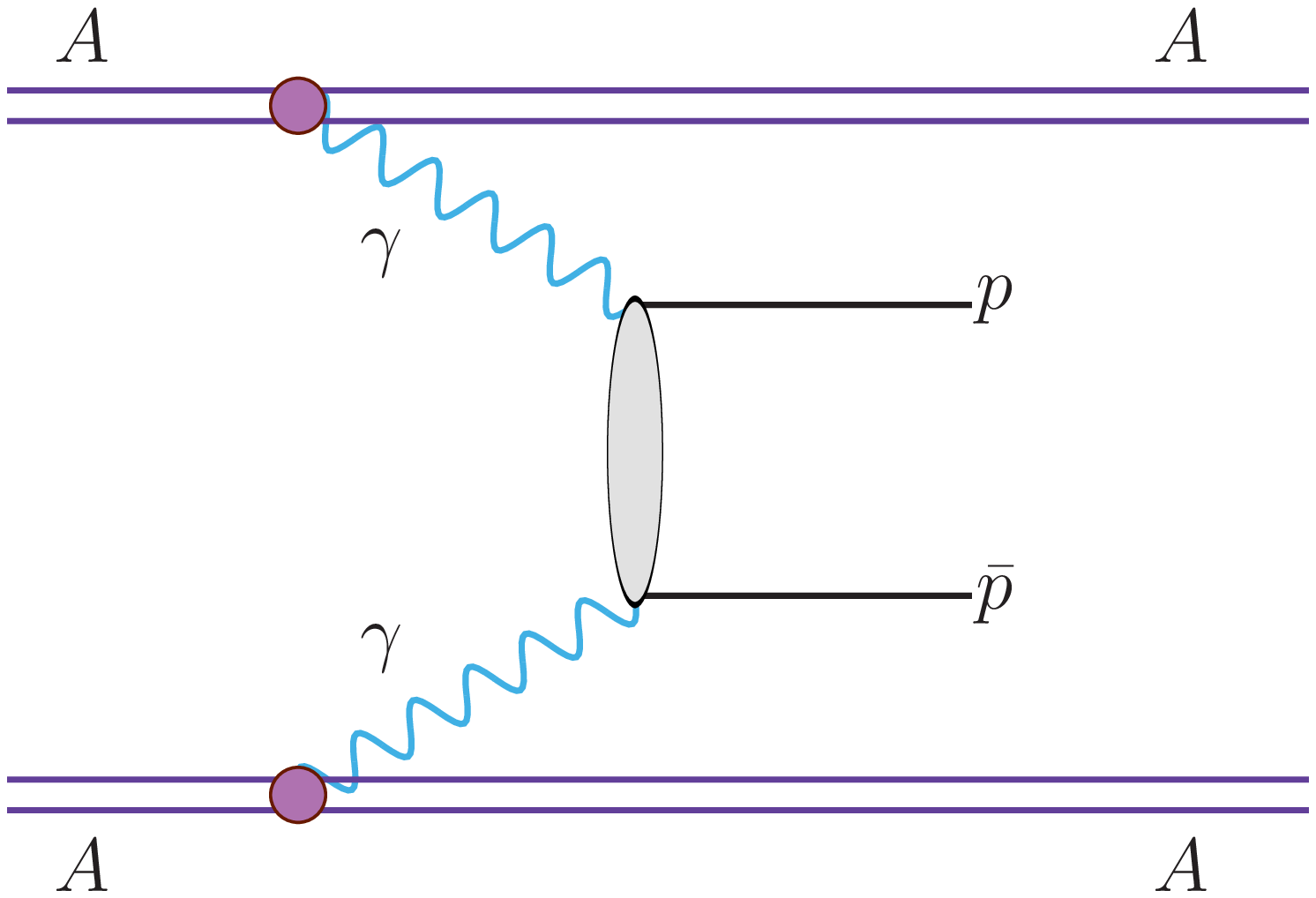}\\
(b)\includegraphics[width=4.5cm]{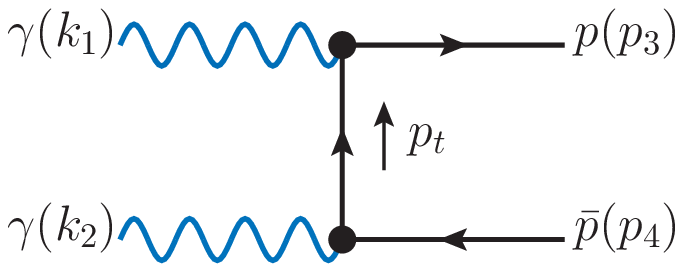}
(c)\includegraphics[width=4.5cm]{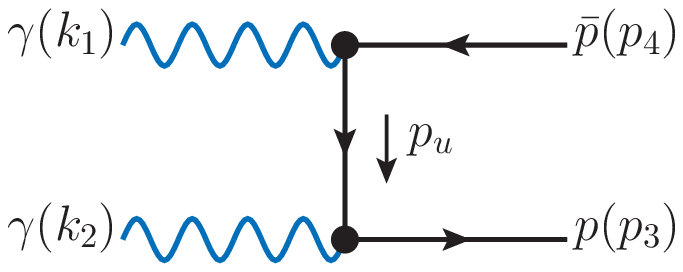}\\
(d)\includegraphics[width=4.5cm]{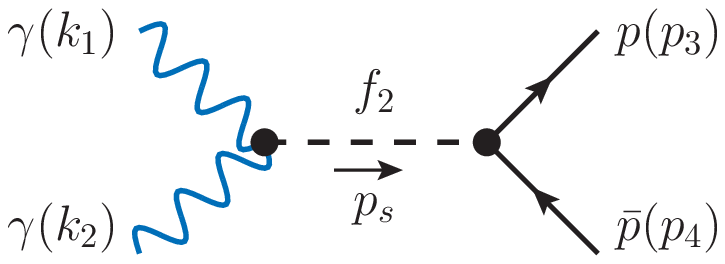}
(e)\includegraphics[width=5.5cm]{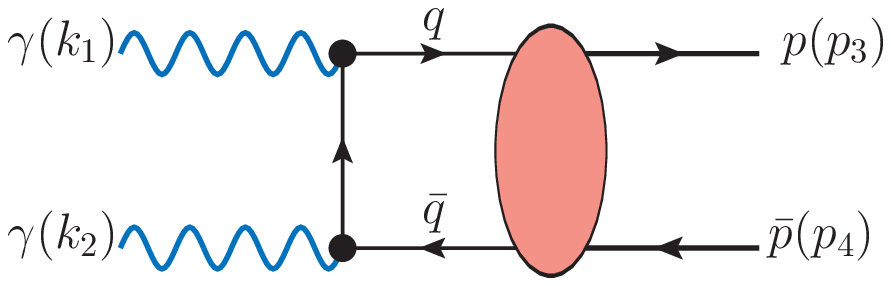}
\caption{Diagram (a) represents $p \bar{p}$ production in ultrarelativistic 
ultraperipheral collisions (UPC) of heavy ions
and other diagrams describe the $\gamma \gamma \to p \bar{p}$ subprocess;
the $t$- and $u$-channel proton exchange
(diagrams (b) and (c), respectively),
the exchange of $f_{2}$ meson in the $s$-channel (diagram (d))
and the hand-bag mechanism (diagram (e)
plus the one with the photon vertices interchanged).
}
\label{fig:AA_AAppbar}
\end{figure}
%--------------------------------------------------------

We focus on the process for ultraperipheral collisions of heavy ions
\begin{eqnarray}
^{208}\!Pb + ^{208}\!\!Pb \to ^{208}\!\!Pb + ^{208}\!\!Pb + p + \bar{p}\,,
\label{nuclear_reaction}
\end{eqnarray}
see diagram (a) shown in Fig.~\ref{fig:AA_AAppbar}.
The nuclear cross section is calculated in the equivalent photon approximation
in the impact parameter space $b = |\bb|$;
for more details see \cite{Klusek-Gawenda:2017lgt}.
%This approach allows to take into account the transverse distance between the colliding nuclei.
The total (phase space integrated) cross section 
is expressed through the five-fold integral
\begin{eqnarray}
\sigma_{AA \to AA p\bar{p}} \left( \sqrt{s_{AA}} \right)&=& 
\int \sigma_{\gamma\gamma \to p\bar{p}}(W_{\gamma\gamma})  
N(\omega_1, \bba) N(\omega_2, \bbb) 
S^2_{abs}(\bb) \nonumber \\
&& \times  \frac{W_{\gamma\gamma}}{2}
dW_{\gamma \gamma} \, d{\rm Y}_{p\bar{p}} \, d\overline{b}_x \, d\overline{b}_y 
\, 2 \pi \,b\,db \,,
\label{eq:sig_nucl_tot}
\end{eqnarray}  
where the impact parameter $b$ means the distance between colliding nuclei
in the plane perpendicular to their direction of motion,
$W_{\gamma\gamma}=\sqrt{4\omega_1\omega_2}$ is the invariant
mass of the $\gamma\gamma$ system, and $\omega_i$, $i=1,2$,
is the energy of the photon which is emitted from the first or second nucleus, respectively.
${\rm Y}_{p\bar{p}}=\frac{1}{2}({\rm y}_{p} + {\rm y}_{\bar{p}})$
is the rapidity of the $p\bar{p}$ system. 
The quantities 
$\overline{b}_x = (b_{1x}+b_{2x})/2$, $\overline{b}_y=(b_{1y}+b_{2y})/2$
are given in terms of $b_{ix}$, $b_{iy}$ which are the components of 
the $\bba$ and $\bbb$ vectors 
which mark a point (distance from first and second nucleus) 
where photons collide and particles are produced. 
%The diagram illustrating these quantities in the impact parameter
%space can be found in \cite{KlusekGawenda:2010kx}. 
In Ref.~\cite{KlusekGawenda:2010kx} the dependence of the photon flux    
$N\left( \omega_i, \bbi \right)$ on the charge form factors of the colliding nuclei was shown explicitly.
In our calculations we use the so-called realistic form factor 
which is the Fourier transform of the charge distribution in the nucleus.
%A more detailed discussion of this issue is given in~\cite{KlusekGawenda:2010kx}.
The presence of the absorption factor $S^{2}_{abs}(\bb)$ in Eq.~(\ref{eq:sig_nucl_tot})
assures that we consider only peripheral collisions, 
when the nuclei do not undergo nuclear breakup.
%In the first approximation this geometrical factor
%can be expressed as
%$S^{2}_{abs}(\bb) = \theta(|\bb|-(R_{A}+R_{B})) = \theta(|\bba-\bbb|-(R_{A}+R_{B}))$
%where the sum of the radii of the two nuclei occurs.

\section{Results for the $\gamma \gamma \to p \bar{p}$ reaction}

%--------------------------------------------------------
\begin{figure}[htb]
\includegraphics[width=6.4cm]{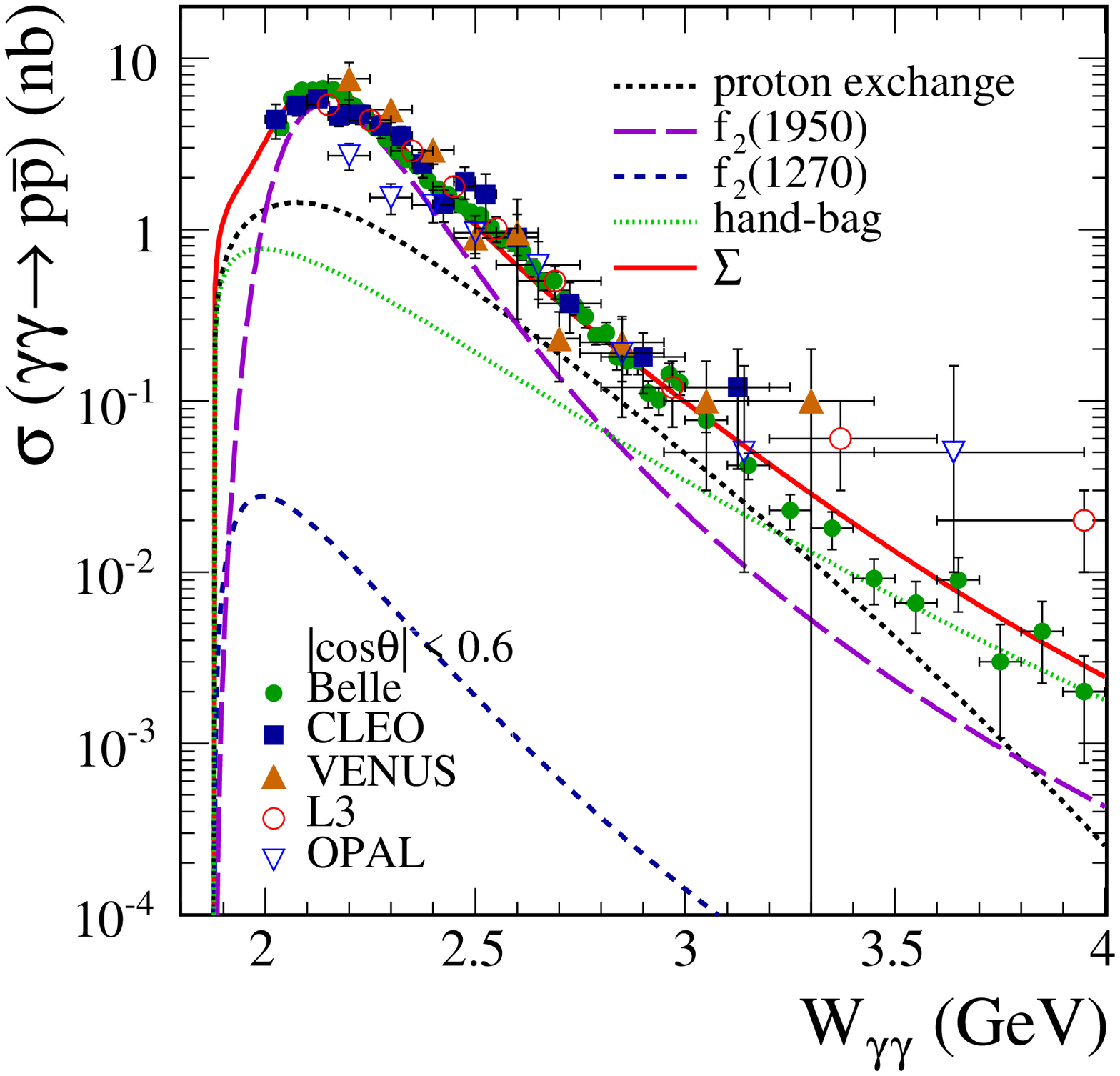}
\includegraphics[width=6.3cm]{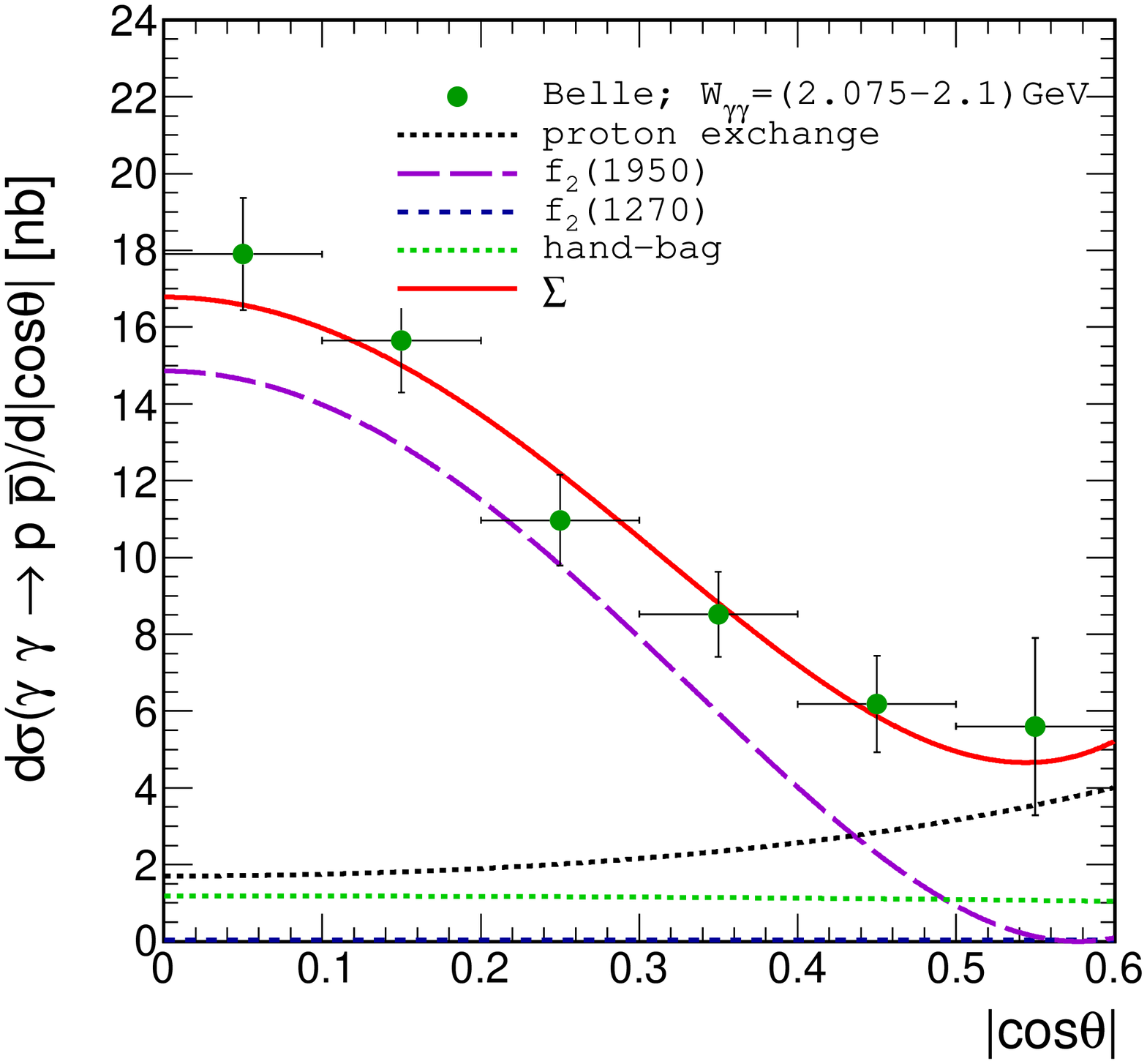}\\
\includegraphics[width=6.3cm]{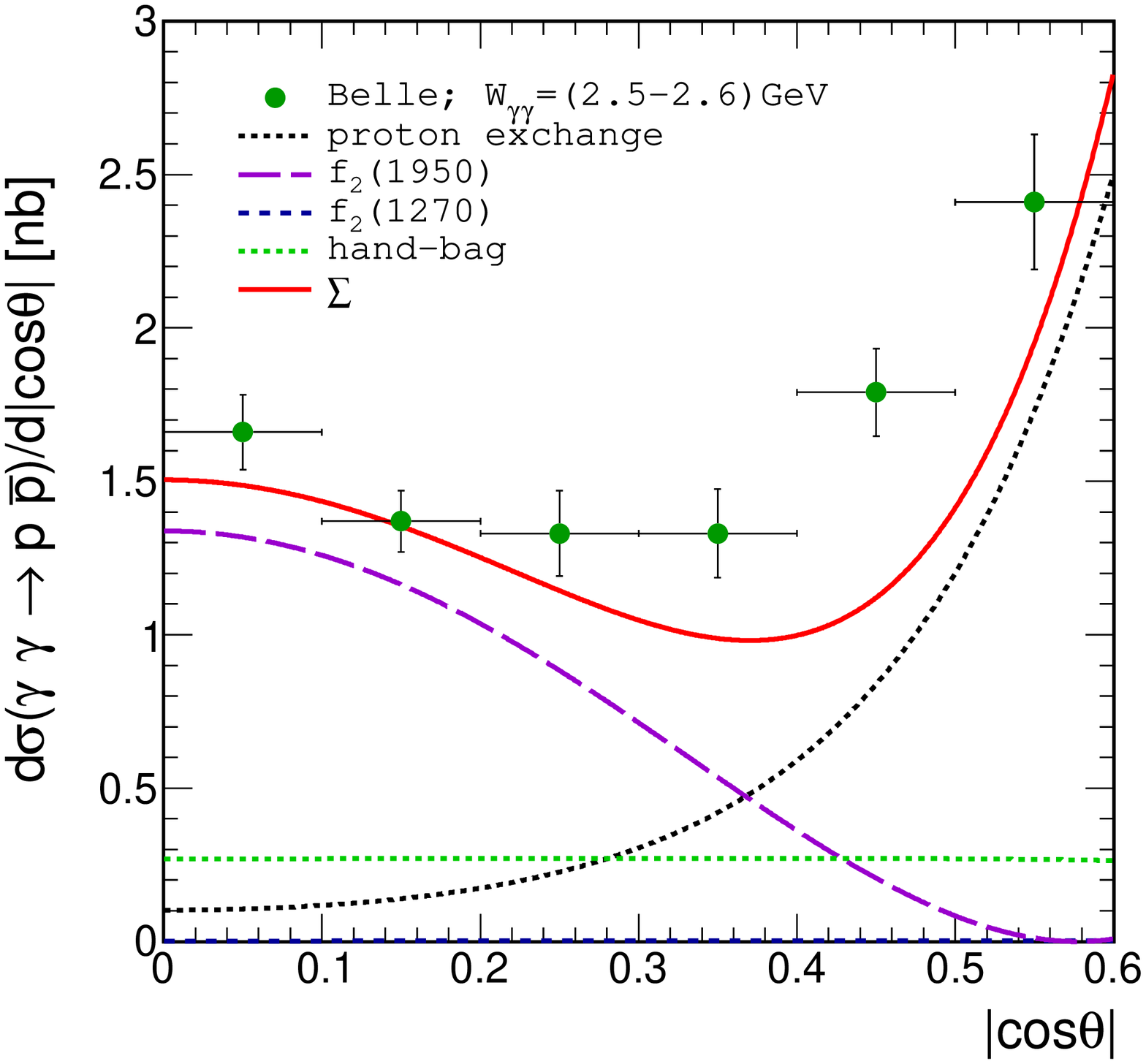}
\includegraphics[width=6.3cm]{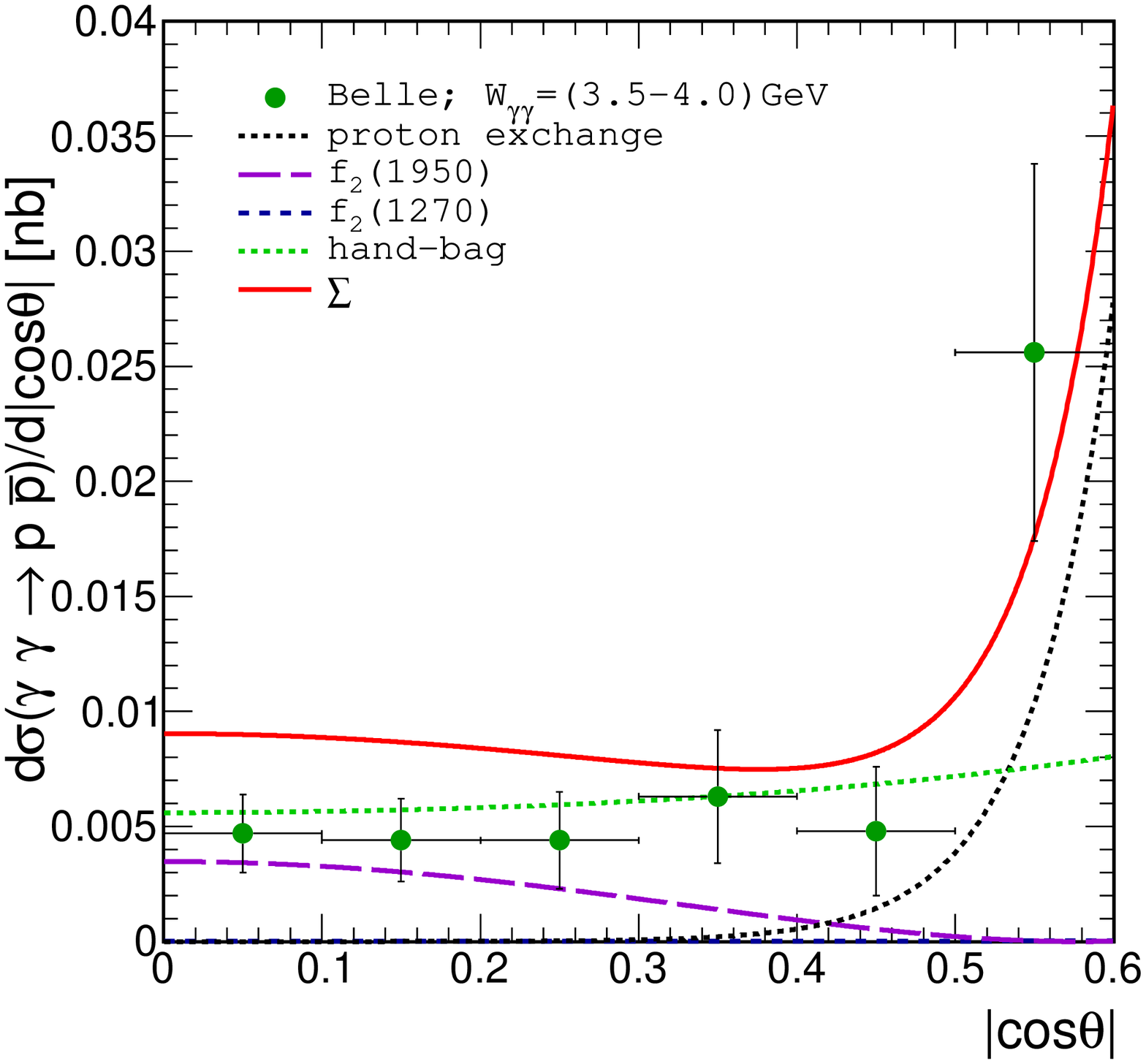}
\caption{Energy dependence of the total cross section 
for $\gamma \gamma \to p \bar{p}$ for $|\cos \theta|<0.6$
and our fit to the Belle angular distributions.
Here, the theoretical results for the parameter set~B 
from Table~II of \cite{Klusek-Gawenda:2017lgt} are shown.
The experimental data are from the CLEO \cite{Artuso:1993xk},
VENUS \cite{Hamasaki:1997cy}, OPAL \cite{Abbiendi:2002bxa},
L3 \cite{Achard:2003jc}, and Belle \cite{Kuo:2005nr} experiments.}
\label{fig:sigma_W_3mech}
\end{figure}
%--------------------------------------------------------
In Fig.~\ref{fig:sigma_W_3mech} we show the energy dependence of the cross section 
for the $\gamma \gamma \to p \bar{p}$ reaction together with the experimental data.
In the Belle experiment \cite{Kuo:2005nr} the $\gamma \gamma \to p \bar{p}$ cross sections
were extracted from the $e^+ e^- \to e^+ e^- p \bar p$ reaction
for the $\gamma \gamma$ c.m. energy range of $2.025 < W_{\gamma\gamma} < 4$~GeV
and in the c.m. angular range of $|\cos\theta|<0.6$.
In Fig.~\ref{fig:sigma_W_3mech} we show also our fit to the Belle angular distributions 
for the three selected intervals of $W_{\gamma\gamma}$.
We take into account the nonresonant proton exchange contribution, 
the $s$-channel tensor meson exchange contributions
and the hand-bag mechanism, see the diagrams in Fig.~\ref{fig:AA_AAppbar} (b) -(e).
Here, the results for the parameter set~B from Table~II of \cite{Klusek-Gawenda:2017lgt} are presented.
One can observe the dominance of the $f_{2}(1950)$ resonance term at low energies.
The proton exchange contribution plays an important role from the threshold
to higher energy while the hand-bag contribution only at $W_{\gamma\gamma} > 3$~GeV.
In our calculation of the nonresonant proton exchange
we have included both Dirac- and Pauli-type couplings of the photon to the nucleon
and form factors for the exchanged off-shell protons.
We have found that the Pauli-type coupling is very important, 
enhances the cross section considerably, and cannot therefore be neglected.

\section{Predictions for the nuclear ultraperipheral collisions}
%--------------------------------------------------------
%\begin{figure}[htb]
%\centerline{
%\includegraphics[width=8cm]{paw_zW_nucl}}
%\caption{Distribution in ($z,W_{\gamma\gamma}$)
%for the $PbPb \to Pb Pb p \bar{p}$ reaction at $\sqrt{s_{NN}}=5.02$ TeV.}
%\label{fig:nuclear_dsig_dzdW}
%\end{figure}
%--------------------------------------------------------

%--------------------------------------------------------
\begin{figure}[htb]
\centerline{
\includegraphics[width=6.5cm]{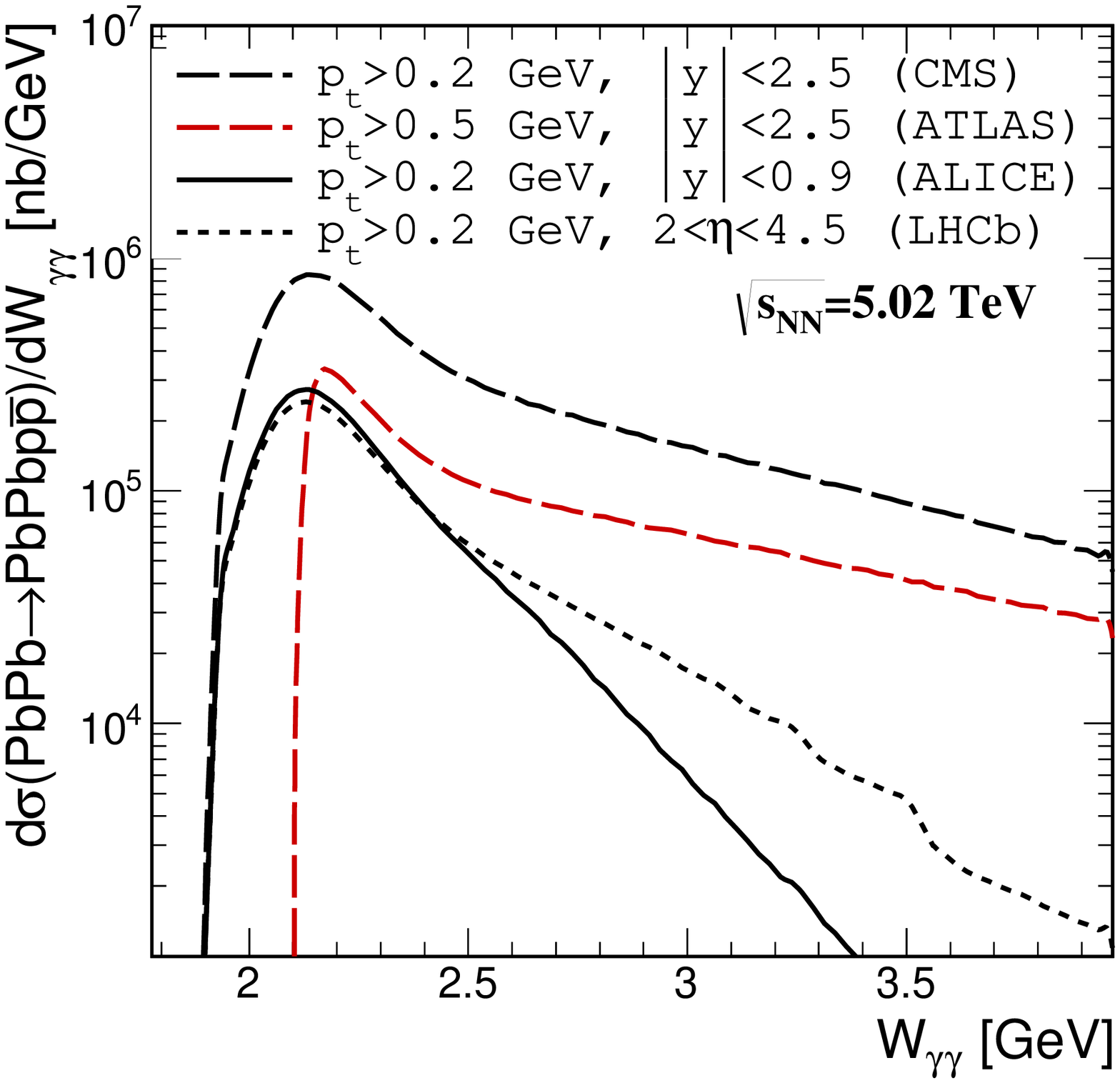}
\includegraphics[width=6.5cm]{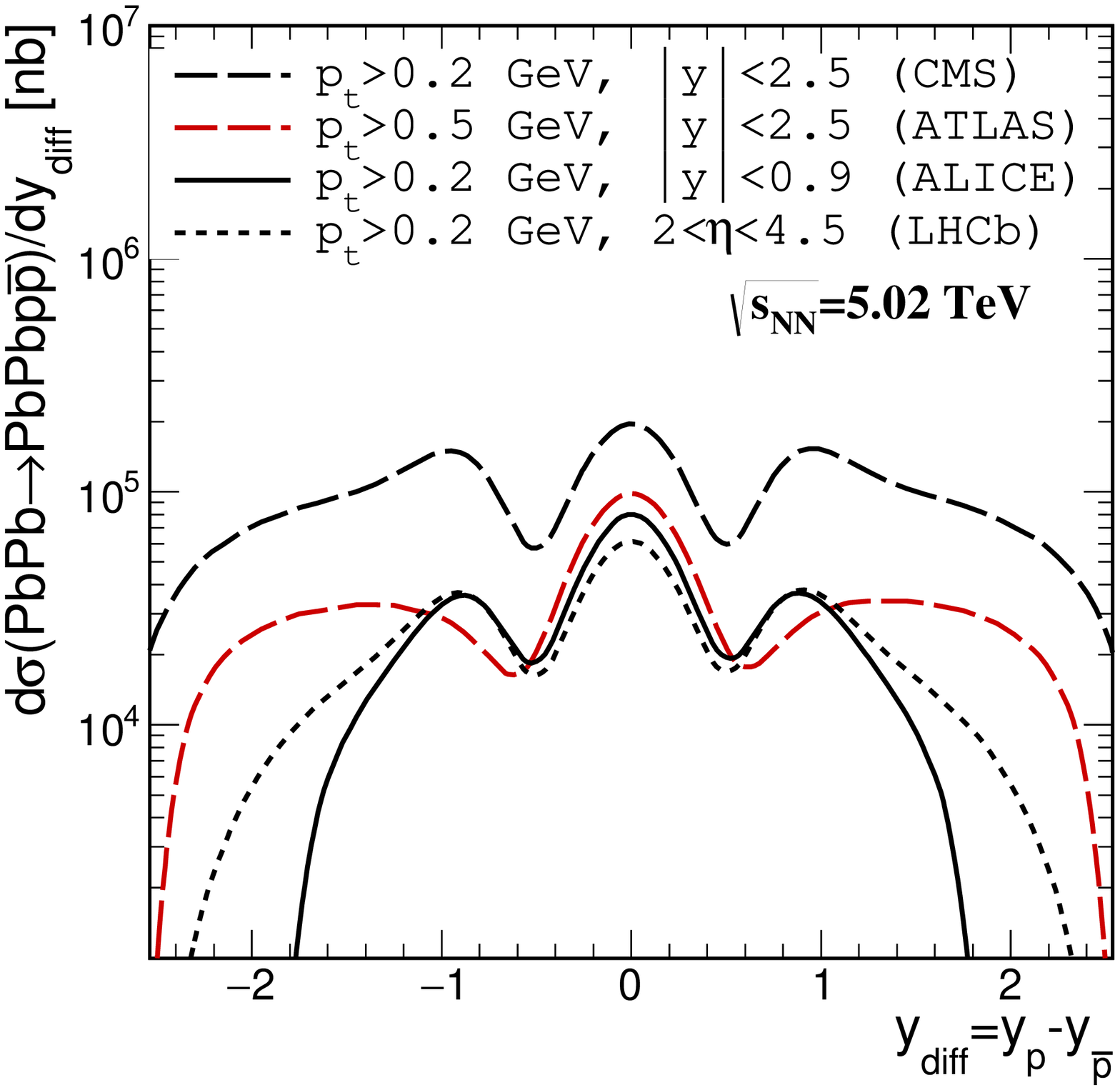}}
\caption{The differential nuclear cross sections as a function of $p \bar{p}$ invariant mass
(the left panel) and ${\rm y}_{diff}={\rm y}_p-{\rm y}_{\bar{p}}$ (the right panel) 
for the $Pb Pb \to Pb Pb p \bar{p}$ reaction at $\sqrt{s_{NN}} = 5.02$~TeV.
The results for different experimental cuts are presented.}
\label{fig:nuclear_ALICE}
\end{figure}
%--------------------------------------------------------
In Fig.~\ref{fig:nuclear_ALICE} we present distributions in 
$W_{\gamma \gamma} \equiv M_{p \bar{p}}$ (the left panel) and 
$\mathrm{y}_{diff} = \mathrm{y}_{p} - \mathrm{y}_{\bar{p}}$ (the right panel) 
imposing cuts on rapidities and transverse momenta of outgoing baryons.
%The upper lines relate to a $|{\rm y}|<2.5$ limit while 
%the lower ones correspond to a narrower limit: $|{\rm y}|<0.9$.
From the left panel, we can observe that
the dependence on invariant mass of the $p \bar{p}$ pair 
is sensitive to the (pseudo)rapidity cut imposed.
%Note that due to the cut on $p_t > 0.5$~GeV
%the $W_{\gamma \gamma}$ distribution begins with a larger value of 2.1~GeV
%(compare also with Fig.~\ref{fig:nuclear}~(a)).
From Figs.~12 - 14 of \cite{Klusek-Gawenda:2017lgt} we clearly see that
results for the nuclear reaction correspond to 
that for elementary $\gamma\gamma \to p \bar{p}$ reaction.
The $f_{2}(1950)$ contribution dominates at smaller $W_{\gamma\gamma}$ and
at $z \approx 0$ and $z \approx \pm 1$
($z = \cos\theta$ in the $\gamma \gamma$ c.m. system). 
This coincides with the result which was presented 
in Fig.~\ref{fig:sigma_W_3mech}, see Fig.~6 of \cite{Klusek-Gawenda:2017lgt}.
In contrast to the resonant contribution, 
the proton-exchange one is concentrated mostly at larger invariant masses
and around $z = \pm 1$.   
The cross section is concentrated along the diagonal ${\rm y}_p \simeq {\rm y}_{\bar{p}}$.
The distribution in the difference of proton and antiproton rapidities is interesting.
%Again (comparing with Fig.~\ref{fig:nuclear}~(d), $|z|<1.0$)
The larger the range of phase space
the broader is the $\mathrm{y}_{diff}$, i.e.,
the larger rapidity distance between $p$ and $\bar{p}$.
There three maxima are visible.
% and the experimental cuts imposed on $p_{t}$ 
%do not remove the external maxima predicted by our model.
The broad peak at $\mathrm{y}_{diff} \approx 0$ corresponds to the region $|z| < 0.6$
which for low-$M_{p \bar{p}}$ is dominated by the $f_{2}(1950)$ term.
It seems that observation of the broader $\mathrm{y}_{diff}$ distribution,
in particular identification of the outer maxima, could be a good test of model.

\section{Conclusions}

We have discussed the production of proton-antiproton pairs in photon-photon interactions.
We have shown that the Belle data \cite{Kuo:2005nr} for low photon-photon energies
can be nicely described by including in addition to the proton exchange 
the $s$-channel exchange of the $f_2(1950)$ resonance
which was observed to decay into the $\gamma\gamma$ and $p \bar{p}$ channels.
%and the hand-bag mechanism. 
%We include in the calculation also the $s$-channel $f_2(1270)$ meson exchange contribution.
%These two tensor mesons were also needed to describe the Belle data 
%for the $\gamma\gamma \to \pi^+\pi^-$ and $\gamma\gamma \to \pi^0\pi^0$
%processes \cite{Uehara:2009cka, Klusek-Gawenda:2013rtu}.
%Our simple model has a few parameters; see Table~\ref{table:parameters}.
Adjusting the parameters of the vertex form factors for the proton exchange,
of the tensor meson $s$-channel exchanges,
and the parameters in the hand-bag contribution 
we have managed to describe both total cross section
and differential angular distributions of the Belle Collaboration.

Having described the Belle data we have used the
$\gamma \gamma \to p \bar p$ cross section to calculate 
the predictions for the
$^{208}\!Pb \,^{208}\!Pb \to \,^{208}\!Pb \,^{208}\! Pb \,\,p \bar{p}$ reaction
at $\sqrt{s_{NN}} = 5.02$~TeV with the LHC experimental cuts.
Large cross sections of 0.1 - 0.5~mb have been obtained. 
We have presented distributions in the invariant mass of the $p \bar p$ system as well as
in the difference of rapidities for protons and antiprotons.
The UPC of heavy ions may provide 
new information compared to the presently available data from
$e^+ e^-$ collisions, in particular, when the structures
of the $\mathrm{y}_{diff}$ distribution can be observed.
%The larger the range of phase space
%the broader is the distribution in $\mathrm{y}_{diff}$,
%the rapidity difference between proton and antiproton.
%We find the cross section of 100~$\mu$b taking into account the ALICE cuts 
%($|{\rm y}|<0.9$, $p_{t}>0.2$~GeV),
%160~$\mu$b for the ATLAS cuts ($|{\rm y}|<2.5$, $p_{t}>0.5$~GeV),
%500~$\mu$b for the CMS cuts ($|{\rm y}|<2.5$, $p_{t}>0.2$~GeV),
%and 104~$\mu$b for the LHCb cuts ($2<\eta<4.5$, $p_{t}>0.2$~GeV).

%--------------------
%\begin{center}
%{\bf Acknowledgments}\\
%\end{center}
%--------------------
%Thanks Mariola K{\l}usek-Gawenda, Antoni Szczurek and Otto Nachtmann
%for cooperation.
This work was supported by the National Science Centre, Poland (NCN)
(grant number 2014/15/B/ST2/02528).


\begin{thebibliography}{1000}

\bibitem{Klusek-Gawenda:2017lgt}
M. K{\l}usek-Gawenda, P. Lebiedowicz, O. Nachtmann, A. Szczurek, \textit{Phys. Rev. D} {\bf 96}, 094029 (2017).

\bibitem{Artuso:1993xk}
M. Artuso \textit{et al.} [CLEO Collaboration], \textit{Phys. Rev. D} {\bf 50}, 5484 (1994).

\bibitem{Hamasaki:1997cy}
H. Hamasaki \textit{et al.} [VENUS Collaboration], \textit{Phys. Lett. B} {\bf 407}, 185 (1997).

\bibitem{Abbiendi:2002bxa}
G. Abbiendi \textit{et al.} [OPAL Collaboration], \textit{Eur. Phys. J. C} {\bf 28}, 45 (2003).

\bibitem{Achard:2003jc}
P. Achard \textit{et al.} [L3 Collaboration], \textit{Phys. Lett. B} {\bf 571}, 11 (2003).

\bibitem{Kuo:2005nr}
C.C. Kuo \textit{et al.} [Belle Collaboration], \textit{Phys. Lett. B} {\bf 621}, 41 (2005).

\bibitem{Chernyak:1984bm}
V.L. Chernyak and I.R. Zhitnitsky, \textit{Nucl. Phys. B} {\bf 246}, 52 (1984).

\bibitem{Farrar:1988vz}
G.R. Farrar, H. Zhang, A.A. Ogloblin, I.R. Zhitnitsky, \textit{Nucl. Phys. B} {\bf 311}, 585 (1989).

\bibitem{Berger:2002vc}
C.F. Berger and W. Schweiger, \textit{Eur. Phys. J. C} {\bf 28} ,249 (2003).

\bibitem{Diehl:2002yh}
M. Diehl, P. Kroll, C. Vogt, \textit{Eur. Phys. J. C} {\bf 26}, 567 (2003).

\bibitem{Odagiri:2004mn}
K. Odagiri, \textit{Nucl. Phys. A} {\bf 748}, 168 (2005).

\bibitem{Lebiedowicz:2018sdt}
P. Lebiedowicz, O. Nachtmann, A. Szczurek, \textit{Phys. Rev. D} {\bf 97}, 094027 (2018).

\bibitem{KlusekGawenda:2010kx}
M. K{\l}usek-Gawenda and A. Szczurek, \textit{Phys. Rev. C} {\bf 82}, 014904 (2010).

\end{thebibliography}
\end{document}